\documentclass[twocolumn,prl,superscriptaddress,showpacs]{revtex4}
\usepackage{epsf,graphicx,amssymb}
\begin{document}
\draft
\title{Chaotic magnetic field reversals in turbulent dynamos}

\author{Christophe Gissinger}
\affiliation{Laboratoire de Physique Statistique de l'Ecole Normale Sup\'erieure, CNRS UMR 8550, 24 Rue Lhomond, 75231 Paris Cedex 05, France}
\author{Emmanuel Dormy}
\affiliation{MAG (IPGP/ENS), CNRS UMR 7154, LRA, Ecole Normale Sup\'erieure, 24 Rue Lhomond, 75231 Paris Cedex 05, France}
\author{Stephan Fauve}
\affiliation{Laboratoire de Physique Statistique de l'Ecole Normale
Sup\'erieure, CNRS UMR 8550, 24 Rue Lhomond, 75231 Paris Cedex 05, France}

\def\bfnabla{\mbox{\boldmath $\nabla$}}
\def\bfv{\mbox{\boldmath $v$}}
\def\bfF{\mbox{\boldmath $F$}}
\def\bff{\mbox{\boldmath $f$}}

\date{\today}
\begin{abstract}
We present direct numerical simulations of reversals of the magnetic field generated by swirling flows in a spherical domain. In agreement with a recent model, we observe that coupling dipolar and quadrupolar magnetic modes by an asymmetric forcing of the flow generates field reversals. In addition, we show that this mechanism strongly depends on the value of the magnetic Prandtl number.  

\end{abstract}
\maketitle 

The generation of magnetic field by the flow of an electrically conducting fluid, i.e., the dynamo effect, has been mostly studied to understand the magnetic fields of planets and stars~\cite{moffatt78}. The Earth and the Sun provide the best documented examples: they both involve a spatially coherent large scale component of magnetic field with well characterized dynamics. Earth's dipole is nearly stationary on time scales much larger than the ones related to the flow in the liquid core, but displays random reversals. Reversals also occur for the Sun but nearly periodically. The magnetic field changes polarity roughly every $11$ years.  Reversals have been displayed by direct simulations of the equations of magnetohydrodynamics (MHD)~\cite{simulations} or of mean field MHD~\cite{stefani} and have been modeled using low dimensional dynamical systems~\cite{dst,nozieres}. 
It has been observed recently that the magnetic field generated by a von Karman flow of liquid sodium (VKS experiment) can display either periodic or random reversals~\cite{berhanu07} as well as several other dynamo regimes, all located in a small parameter range~\cite{ravelet08}. 
The ability of all these very different dynamos to reverse polarity is their most striking property. This is obviously related to the ${\bf B} \rightarrow - {\bf B}$ symmetry of the MHD equations, implying that if a magnetic field ${\bf B}$ is a solution,  $-{\bf B}$ is another solution. However, this does not explain how these two solutions can be connected as time evolves. The VKS experiment has provided an interesting observation. In this experiment, the flow is driven in a cylindrical container by two counter-rotating coaxial propellers. When they rotate at roughly the same frequency, a magnetic field with a dominant dipolar component aligned with the axis of rotation is generated. Time dependent magnetic field with periodic or random reversals are observed only when the difference between the two rotation frequencies is large enough~\cite{berhanu07}. We have shown that this can be related to broken ${\cal R}_{\pi}$ symmetry (the rotation of an angle $\pi$ along any axis in the mid-plane is a symmetry when the propellers rotate at the same frequency)~\cite{petrelis08}. Magnetic modes changed to their opposite by ${\cal R}_{\pi}$ are dipolar ones, whereas quadrupolar ones are unchanged. 
Breaking the ${\cal R}_{\pi}$ symmetry by rotating the two propellers at different frequencies generates additional coupling terms between dipolar and quadrupolar modes. If their dynamo threshold is close enough, a slightly broken symmetry is sufficient to lead to a saddle node bifurcation, which generates a limit cycle that connects opposite polarities.  Slightly below this bifurcation threshold, ${\pm \bf B}$ stationary solutions are stable but a small amount of hydrodynamic fluctuations is enough to generate random reversals. Although the flow in the Earth's core strongly differs from the one of the VKS experiment, a similar type of interaction between two marginal dynamo modes can be considered and provides a simple explanation of several features of paleomagnetic records of Earth's reversals~\cite{petrelis09}. 

The purpose of this work is to strengthen this phenomenological scenario by displaying reversal of a magnetic dipole coupled with a quadrupolar mode in a direct numerical simulation of MHD equations. To wit, we consider a flow driven in a spherical geometry by volumic forces that mimic the motion of two co-axial propellers. We observe reversals of the generated magnetic field for a wide range of parameters when the propellers rotate at different speeds. We show that the value of the magnetic Prandtl number $P_m$ strongly affects the magnetic modes involved in the dynamics of reversals. Reversals that involve a coupling between dipole and quadrupole modes occur for $P_m$ small enough. Finally, we present a minimal model for the reversal dynamics.\\

The MHD equations are integrated in a spherical geometry for the
solenoidal velocity $\bf v$ and magnetic $\bf B$ fields,
\begin{eqnarray}
\frac{\partial \bf{v} }{ \partial t} + (\bf{v}\cdot\bfnabla) \bf{v} \!&=&\! -
\bfnabla \pi + \nu 
\Delta \bf{v} + \bf{f} +\frac{1}{\mu_0 \rho}(\bf {B} \cdot \bfnabla) \bf
{B}, \label{ns}\\
\frac{\partial \bf{B}}{ \partial t} \!&=&\! \bfnabla
\times \left(\bf{v} \times \bf{B}\right) + \frac{1}{\mu_0 \sigma} \, 
\Delta \bf {B}. \label{ind}
\end{eqnarray}
In the above equations, $\rho$ is the density, $\mu_0$ is the magnetic
permeability and $\sigma$ is the electrical conductivity of the fluid.  The forcing is ${\bf f} = f_0\,{\bf F}$, where $F_\phi=s^2 \sin(\pi\, s
\,b)\, , \; F_z=\varepsilon\, \sin(\pi \, s \, c) \, ,$ for $z>0$, and opposite for $z<0$. We use polar coordinates $(s, \phi, z)$, normalized by the radius of the sphere $a$.   $F_\phi$ generates counter-rotating flows in each hemisphere, while $F_z$ enforces a strong poloidal circulation. The forcing is only applied in the region $0.25a<\mid\!z\!\mid<0.65a$, $s<s_0$.  In the simulations presented
here, $s_0=0.4$, $b^{-1}=2s_0$ and $c^{-1}=s_0$.  This forcing has
been used to model the Madison experiment~\cite{bayliss07}. It is invariant by the ${\cal R}_{\pi}$ symmetry. In order to break it, we consider in the present study a forcing of the form
${C\bf f}$, where $C=1$ for $z<0$ but can be different from one for $z>0$. This describes two propellers that counter-rotate at different frequencies.  Although performed in a spherical geometry, this simulation involves a mean flow with a similar topology to that of the VKS experiment.  We solve the above system of equations using the Parody numerical code~\cite{parody}.  This code was originally developped in the context of the geodynamo (spherical shell) and we have modified it to make it suitable for a full sphere.  We use the same dimensionless numbers as in~\cite{gissinger08}, the magnetic Reynolds number $R_m= \mu_0\sigma a \, {\rm max} (\vert \bf v \vert)$, and the magnetic Prandtl number $P_m=\nu\mu_0\sigma$.  The kinetic Reynolds number is then $Re=R_m/P_m$.

\begin{figure}
\centerline{
\epsfxsize=0.3\textwidth 
\epsffile{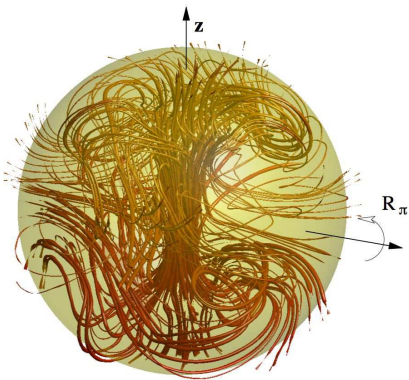} }
\vskip -3mm
\caption{Magnetic field lines obtained with a symmetric forcing ($C=1$) for $R_m=300$ and $P_m=1$. Note that the field involves a dipolar component with its axis aligned with the axis $z$ of rotation of the propellers.}
\label{axialdipole}
\end{figure}

In a previous study with symmetric forcing ($C=1$), we showed that different magnetic modes can be generated depending on $P_m$~\cite{gissinger08}. For large enough $P_m$, the dynamo onset corresponds to small $Re$ and the flow is axisymmetric and generates first an equatorial dipole. In contrast, for $P_m$ small, the dynamo onset occurs when $Re$ is already large and the flow involves non axisymmetric fluctuations.  A magnetic field with a dominant axial dipole is observed (see figure~\ref{axialdipole}). In the present study, all the simulations are made for $Re > 300$, so that an axial dipole is obtained for symmetric forcing. 

\begin{figure}
\centerline{
\epsfxsize=0.5\textwidth 
\epsffile{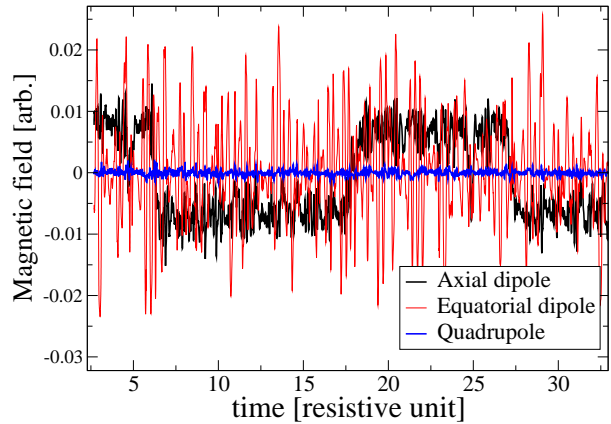} }
\vskip -4mm
\caption{Time recording of the axial dipolar magnetic mode (in black),
the axial quadrupolar mode (in blue) and the equatorial dipole (in red) for $R_m=300$, $P_m=1$ and $C=2.$. }
\label{Pm1}
\end{figure}

We next break the symmetry ${\cal R}_{\pi}$ of the forcing  to check whether time dependent magnetic fields involving reversals between both polarities are obtained as in the VKS experiment. Time recordings of some components of the magnetic field are displayed in figure~\ref{Pm1} for $R_m=300$, $P_m=1$ and $C=2$, which means that one of the propellers is spinning twice as fast as the other one.  We observe that the axial dipolar component (in black) randomly reverses sign. The phases with given polarity are an order of magnitude longer than the duration of a reversal that corresponds to an Ohmic diffusion time. The magnetic field strongly fluctuates during these phases because of hydrodynamic fluctuations. It also displays excursions or aborted reversals, i.e., the dipolar component almost vanishes or even slightly changes sign but then grows again with its direction unchanged. All these features are observed in paleomagnetic records of Earth's magnetic field~\cite{valet05} and also in the VKS experiment~\cite{berhanu07}. However, these simulations also display strong differences with the VKS experiment. The equatorial dipole is the mode with the largest fluctuations whereas the axial quadrupolar components is an order of magnitude smaller than the equatorial modes. In addition, it does not seem to be coupled to the axial dipolar component. 

\begin{figure*}
\centerline{
\epsfxsize=1\textwidth 
\epsffile{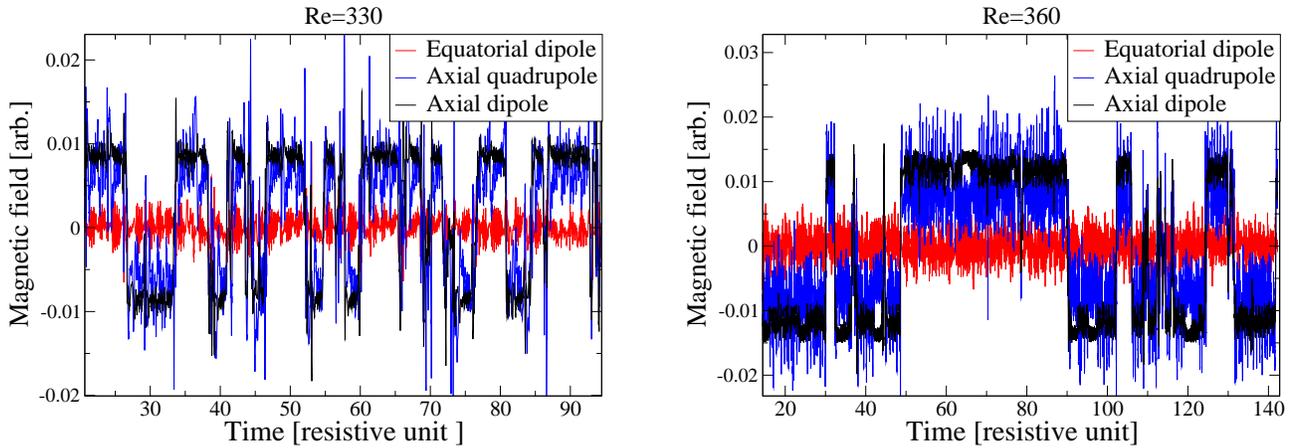} }
\vskip -4mm
\caption{Time recordings of the axial dipole (black),
the axial quadrupole (blue) and the equatorial dipole (red). Left: $R_m=165$, $P_m=0.5$ and $C=1.5$. Right: $R_m=180$, $P_m=0.5$ and $C=2$.}
\label{Pm0p5}
\end{figure*}

Magnetic Prandtl numbers relevant to liquid metals are much smaller than unity ($\sim 10^{-5}$--$10^{-6}$). While realistic values cannot be achieved owing to computational limitations, $P_m$ can be decreased to values less than unity. We now turn to simulations using $P_m=0.5$, thus introducing a distinction between the viscous and ohmic timescale.  The time evolution of the magnetic modes for $R_m=165$ and $C=1.5$ is represented on figure~\ref{Pm0p5} (left). It differs significantly from the previous case ($P_m=1$). First of all, the quadrupole is now a significant part of the field, and reverses together with the axial dipole. The equatorial dipole remains compartively very weak and unessential to the dynamics. The high amount of fluctuations observed in these signals points to the role of hydrodynamic fluctuations on the reversal mechanism. One could be tempted to speculate that a higher degree of hydrodynamic fluctuations necessarily yields a larger reversal rate. Such is in fact not the case. A more sensible approach could be to try to relate the amount of fluctuations of the magnetic modes in a phase with given polarity, to the frequency of reversals. Increasing $R_m$ from $165$ to $180$  does yield larger fluctuations as shown in figure~\ref{Pm0p5} (right). However the reversal rate is in fact lowered because $C$ was modified to $C=2$. This clearly shows that the asymmetry parameter $C$ plays a more important role than the fluctuations of the magnetic field. For $P_m=0.5$, reversals occur only in a restricted region, $1.1 < C < 2.5$, which is also a feature of the VKS experiment.

\begin{figure}
\centerline{
\epsfxsize=0.45\textwidth 
\epsffile{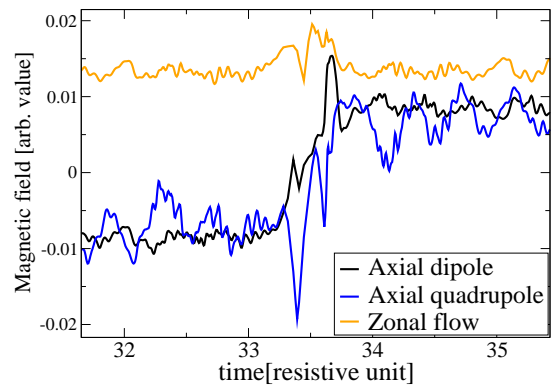} }
\vskip -4mm
\caption{Time recordings of the axial dipole (black),
the axial quadrupole (blue) and zonal velocity (yellow) during a
reversal. $R_m=165$, $P_m=0.5$ and $C=1.5$.}
\label{rev}
\end{figure}

Let us now investigate the detail of a polarity reversal (figure~\ref{rev}). Interestingly the dipolar and quadrupolar components do not vanish simultaneously. Instead the decrease of the dipole is associated with a sudden increase of the quadrupolar component, related with a burst of activity in the non axisymmetric velocity mode $m=1$ (zonal velocity) breaking the ${\cal R}_{\pi}$ symmetry (i.e. coupling the dipolar and quadrupolar families). The quadrupole quickly decays as the dipole recovers with a reversed polarity. Immediatly after the reversal the dipole systematically overshoots its mean value during a polarity interval. This behavior of the magnetic modes is typical of reversals obtained with this value of $P_m$ and is in agreement with the model presented in~\cite{petrelis08,petrelis09}. 

These direct numerical simulations illustrate the role of the magnetic Prandtl number in the dynamics of reversals. When $P_m$ is of order one, the magnetic perturbations due to the advection of magnetic field lines by the velocity field, evolves with a time scale similar to the one of the velocity fluctuations. We thus expect these two fields to be strongly coupled. Modification of the magnetic field lines due to their advection by a local fluctuation of the flow can then trigger a reversal of the field~\cite{parker}. This type of scenario has been observed in some direct numerical simulations, usually performed with $P_m$ of order one~\cite{advection}. When $P_m$ is small, magnetic perturbations decay much faster and we expect only the largest scale magnetic modes to govern the dynamics. To illustrate this argument in a more quantitative way, we have computed the correlation $r$ of the most significant magnetic and velocity modes with the axial magnetic dipole for $Re = 330$ and $0.3 < P_m < 1$. For $P_m \sim 1$, all the modes are weakly correlated with the axial magnetic dipole ($r < 0.3$). When $P_m$ is decreased, the correlation of most modes decay except the one of the zonal velocity mode that slightly increases and the one of the axial quadrupole mode that strongly increases up to $0.8$. 
\begin{figure}
\centerline{
\epsfxsize=0.45\textwidth 
\epsffile{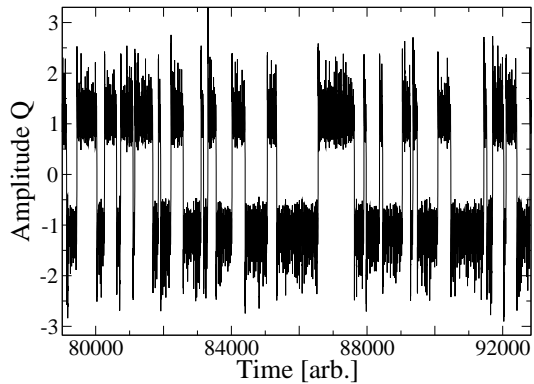} }
\vskip -3mm
\caption{Numerical integration of the amplitude equations~(\ref{ampD},\ref{ampQ},\ref{ampV}). Time recording of the
amplitude of the quadrupolar mode for $\mu = 0.119$, $\nu=0.1$ and $\Gamma = 0.9$.}
\label{model} 
\end{figure}

We now write the simplest dynamical system that involves the three modes that display correlation in the low $P_m$ simulations: the dipole $D$, the quadrupole $Q$, and the zonal velocity mode $V$ that breaks the ${\cal R}_{\pi}$ symmetry. These modes transform as $D \rightarrow - D$, $Q \rightarrow Q$ and $V \rightarrow - V$ under the ${\cal R}_{\pi}$ symmetry. Keeping nonlinear terms up to quadratic order, we get
\begin{eqnarray}
\dot{D} &=& \mu D - V Q, \label{ampD}  \\
\dot{Q} &=& - \nu Q + V D, \label{ampQ} \\
\dot{V} &=& \Gamma - V + Q D. \label{ampV}
\end{eqnarray}
A non zero value of $\Gamma$ is related to a forcing that breaks the ${\cal R}_{\pi}$ symmetry, i.e. propellers rotating at different speeds.
The dynamical system~(\ref{ampD},\ref{ampQ},\ref{ampV}) with $\Gamma = 0$ occurs in different hydrodynamic problems and has been analyzed in detail~\cite{hughes}. The relative signs of the coefficients of the nonlinear terms are such that the solutions do not diverge when $\mu > 0$ and $\nu < 0$. Their modulus can be taken equal to one by appropriate scalings of the amplitudes. The velocity mode is linearly damped and its coefficient can be taken equal to $-1$ by an appropriate choice of the time scale. Note that similar equations were obtained with a drastic truncation of the linear modes of MHD equations~\cite{nozieres}. However, in that context $\mu$ and $\nu$ should be both negative and the damping of the velocity mode was discarded, thus modifying the dynamics.

This system displays reversals of the magnetic modes $D$ and $Q$ for a wide range of parameters. A time recording is shown in figure~\ref{model}. The mechanism for these reversals results from the interaction of the modes $D$ and $Q$ coupled by the broken ${\cal R}_{\pi}$ symmetry when $V \neq 0$. It is thus similar to the one described in~\cite{petrelis08} but keeping the damped velocity mode into the system generates chaotic fluctuations. Thus, it is not necessary to add external noise to obtain random reversals. We do not claim that this minimal low order system fully describes the direct simulations presented here. For instance, in the case of exact counter-rotation ($C = 1$, i.e. $\Gamma = 0$), equations (\ref{ampD},\ref{ampQ},\ref{ampV}) do not have a stable stationary state with a dominant axial dipole. The different solutions obtained when $\mu$ is increased cannot capture all the dynamo regimes of the VKS experiment or of the direct simulations when $R_m$ is increased away from the threshold. Taking into account cubic nonlinearities provides a better description of the numerical results for $P_m = 0.5$. However, this three mode system with only quadratic nonlinearities involves the basic ingredients of the reversals observed in the present numerical simulations for low enough values of the magnetic Prandtl number.

\begin{acknowledgments}
We thank F. P\'etr\'elis for useful discussions. Computations were performed at CEMAG and IDRIS.
\end{acknowledgments}


\end{document}